\documentclass[aps,prd,eqsecnum,showpacs,superscriptaddress]{revtex4}

\usepackage{latexsym}
\usepackage{amsfonts}
\usepackage{amsmath,amssymb}

\newcommand{\dalm}{\kern1pt\vbox{\hrule height 0.9pt\hbox{\vrule width 0.9pt\hskip 2.5pt\vbox{\vskip 5.5pt}\hskip 3pt\vrule width 0.3pt}\hrule height 0.3pt}\kern1pt}
\newtheorem{definition}{Definition}

\begin{document}

\title{Asymptotically anti-de Sitter spacetimes and conserved quantities \\ in higher curvature gravitational theories}

\author{Naoya Okuyama}
\email{okuyama@gravity.phys.waseda.ac.jp}
\affiliation{Department of Physics, Waseda University, Okubo 3-4-1, Shinjuku, Tokyo 169-8555, Japan}
\author{Jun-ichirou Koga}
\email{koga@gravity.phys.waseda.ac.jp}
\affiliation{Advanced Research Institute for Science and Engineering, Waseda University, Okubo 3-4-1, Shinjuku, Tokyo 169-8555, Japan}

\date{\today}
\preprint{WU-AP/209/05}

\begin{abstract}
We consider $n$-dimensional asymptotically anti-de Sitter spacetimes in higher curvature gravitational theories with $n \geq 4$, by employing the conformal completion technique.
We first argue that a condition on the Ricci tensor should be supplemented to define an asymptotically anti-de Sitter spacetime in higher curvature gravitational theories and propose an alternative definition of an asymptotically anti-de Sitter spacetime.
Based on that definition, we then derive a conservation law of the gravitational field and construct conserved quantities in two classes of higher curvature gravitational theories.
We also show that these conserved quantities satisfy a balance equation in the same sense as in Einstein gravity and that they reproduce the results derived elsewhere.
These conserved quantities are shown to be expressed as an integral of the electric part of the Weyl tensor alone and hence they vanish identically in the pure anti-de Sitter spacetime as in the case of Einstein gravity.
\end{abstract}

\maketitle

\section{Introduction}

Recently, higher dimensional asymptotically anti-de Sitter (AdS) spacetimes have been attracting much attention.
In string theory, much work has been done on the anti-de Sitter / conformal field theory (AdS / CFT) correspondence \cite{AdSCFT}, which argues for duality between gravity in an AdS spacetime and CFT on its boundary.
Also in the context of cosmology, the brane world scenario has provided a new perspective \cite{Brane}, where a higher dimensional isolated system in an AdS background plays an important role \cite{Branecosmology}.
There has been also a renewed interest quite recently in the definition of the mass of a rotating black hole in an AdS spacetime \cite{GibbonsEtAl,DeruelleKatz}.
On the other hand, higher curvature corrections to the standard Einstein gravity arise in an effective theory of string theory \cite{HigherOrderGravity} and in the semiclassical quantum field theory on a curved spacetime \cite{Birrell-Davis}.
It is thus interesting to analyze ``asymptotically AdS'' spacetimes in higher curvature gravitational theories.
Actually, ``asymptotically AdS'' black hole solutions have been found in higher curvature gravitational theories \cite{EGBBH} and conserved quantities at infinity of an ``asymptotically AdS'' spacetime have been analyzed \cite{DeruelleKatzOgushi,energyEGBBH,DeruelleMorisawa}.

However, a precise definition of an asymptotically AdS spacetime has been established so far only in Einstein gravity \cite{Ashtekar-Magnon,Ashtekar-Das} but not in higher curvature gravitational theories.
To define an asymptotically AdS spacetime precisely, one can employ the conformal completion technique, which has the advantage that asymptotic behavior of physical fields is analyzed in a well-defined manner.
To do this, we conformally transform a physical spacetime $( M , g_{ab})$ into an unphysical spacetime $(\hat{M} , \hat{g}_{ab})$, to which the asymptotic infinity belongs, so that we can apply the standard method of differential geometry.
Thus, in this paper, by employing the conformal completion technique, we will consider how an $n$-dimensional asymptotically AdS spacetime with $n \geq 4$ is defined in higher curvature gravitational theories, where the action is written as
\begin{equation}
S = \int d^n x \sqrt{-g}\, \left[ \frac{{\cal L}(g^{ab}, R_{abcd})}{16\pi G_{(n)}} + {\cal L}_{\rm matter} \right].
\end{equation}
The Lagrangian ${\cal L}(g^{ab}, R_{abcd})$ of the gravitational field is assumed to be a scalar function of the physical metric $g^{ab}$ and the Riemann tensor $R_{abcd}$ associated with $g_{a b}$, ${\cal L}_{\rm matter}$ is the Lagrangian of ordinary matter fields minimally coupled to the gravitational field, and $G_{(n)}$ is the $n$-dimensional gravitational constant.

The conformal completion technique provides an intrinsic and background-independent definition of conserved quantities in an asymptotically AdS spacetime in Einstein gravity \cite{Ashtekar-Magnon,Ashtekar-Das}.
In addition, by making use of the Einstein equation, those conserved quantities are shown to satisfy a balance equation, which states that decrease of a conserved quantity coincides with the corresponding flux of matter fields radiated away across infinity.
It will be plausible that these features are possessed by higher curvature gravitational theories also.
On the other hand, however, higher curvature gravitational theories are endowed with intricate features that are absent in Einstein gravity.
The gravitational field equation includes nonlinear terms in the curvature tensors, as well as derivatives of the curvature tensors that result in extra degrees of freedom due to higher curvature corrections.
We thus need to define an asymptotically AdS spacetime in higher curvature gravitational theories so that the definition is compatible with these features of higher curvature gravitational theories.

Then, in Sec.\ \ref{sec:A-AdS}, after reviewing the definition in Einstein gravity, we will propose a definition of an asymptotically AdS spacetime by taking into account these features of higher curvature gravitational theories, where we will impose an asymptotic condition on the Riemann tensor, not only on the Weyl tensor.
The geometrical structure and the behavior of physical fields at the asymptotic infinity  that follow from the proposed definition are also described in this section.
In particular, we will point out that the asymptotic symmetry group is unaltered  by higher curvature corrections.

Based on the definition and its consequences in Sec.\ \ref{sec:A-AdS}, we will construct well-defined conserved quantities 
associated with the asymptotic symmetry group in Sec.\ \ref{sec:ConservedQuantities}.
Since construction of conserved quantities depends on the Lagrangian of gravitational theories, we will focus on two classes of higher curvature gravitational theories in this section.
The conserved quantities in these classes of gravitational theories will be shown to satisfy the balance equation in the same sense as in Einstein gravity. In addition, we will find that these conserved quantities are expressed as an integral of (the leading order of) the electric part of the Weyl tensor, while multiplied by a proportion factor that depends on the Lagrangian.
We will also show that the conserved quantities constructed in this paper coincide with the results derived elsewhere.
Finally, in Sec.\ \ref{sec:SummaryandDiscussion}, we will summarize this paper and discuss related issues.

Although the curvature tensors associated with the unphysical metric $\hat{g}_{a b}$ are employed in Ref.\ \cite{Ashtekar-Magnon,Ashtekar-Das}, we will work in this paper with the curvature tensors associated with {\it the physical metric} $g_{a b}$.
This is because an analysis based on the curvature tensors associated with $\hat{g}_{a b}$ becomes quite complicated due to higher curvature corrections.
On the other hand, we need to work in the unphysical spacetime $( \hat{M} , \hat{g}_{a b} )$, anyway.
We thus map the curvature tensors associated with $g_{a b}$ {\it with all indices subscript} from $M$ onto the interior of $\hat{M}$, based on the diffeomorphism between the physical spacetime and unphysical spacetime, e.g., the Riemann tensor $R_{a b c d}$ associated with $g_{a b}$ with all indices subscript is mapped as a tensor with all indices subscript on the interior of $\hat{M}$.
It is understood also that tensor indices are raised or lowered by the unphysical metric $\hat{g}^{a b}$ or $\hat{g}_{a b}$, unless otherwise stated.

\section{Asymptotically anti-de Sitter spacetime}
\label{sec:A-AdS}

\subsection{Einstein gravity}

Before exploring an asymptotically AdS spacetime in higher curvature gravitational theories, we first briefly review the definition of an asymptotically AdS spacetime in Einstein gravity \cite{Ashtekar-Magnon,Ashtekar-Das}.

An $n$-dimensional spacetime $(M , g_{ab})$ in Einstein gravity is said to be asymptotically AdS if there exists a conformally completed spacetime $(\hat{M} , \hat{g}_{ab})$, i.e., an unphysical spacetime to which the boundary ${\cal I}$ of the spacetime belongs.
The boundary ${\cal I}$ represents the asymptotic infinity of the physical spacetime $(M , g_{ab})$ and is assumed to possess the topology of $S^{n-2}\times {\bf R}$.
The interior of $\hat{M}$ is thus diffeomorphic to $M$ and the unphysical metric $\hat{g}_{ab}$ is related with the physical metric $g_{ab}$ by a conformal transformation
\begin{equation}
\hat{g}_{ab} = \Omega^2 g_{ab},
\end{equation}
where the conformal factor $\Omega$ should satisfy
\begin{equation}
\Omega = 0 , ~~ \text{and} ~~ \hat{\nabla}_a \Omega \neq 0
\end{equation}
everywhere on ${\cal I}$, and $\hat{\nabla}_a$ is the covariant derivative associated with the unphysical metric $\hat{g}_{a b}$.
The physical metric $g_{ab}$ is assumed to satisfy the gravitational field equation, i.e., the Einstein equation with a negative cosmological constant, while the energy-momentum tensor $T_{a b}$ of matter fields is required to fall off as $\Omega^{n - 2}$ so that the flux of the matter fields across ${\cal I}$ is finite. Hence, there exists a smooth tensor $\tau_{a b}$, such that $T_{a b}$ is rewritten as
\begin{equation} 
T_{ab} = \Omega^{n-2} \tau_{ab},
\end{equation} 
in a neighborhood of ${\cal I}$.

Since the Weyl tensor identically vanishes in the pure AdS spacetime, the Weyl tensor in an asymptotically AdS spacetime should vanish asymptotically, and then Ashtekar and Das \cite{Ashtekar-Das} imposed a fall-off condition on the Weyl tensor.
(In four dimensions, the `reflective boundary condition' on the Weyl tensor also is imposed, so that the asymptotic symmetry group at infinity reduces to the four-dimensional AdS group $O(3,2)$ \cite{Ashtekar-Magnon}.)
The rate at which the Weyl tensor falls off was determined from dimensional considerations, and it is described in terms of the Weyl tensor $C_{abcd}$ associated with {\it the physical metric} $g_{ab}$ as
\begin{equation}
C_{abcd} = \Omega^{n-5} K_{abcd},
\end{equation}
in a neighborhood of ${\cal I}$, where $K_{abcd}$ is a smooth tensor \footnote{Although this condition is described in terms of the Weyl tensor $\hat{C}_{abcd}$ associated with the unphysical metric $\hat{g}_{ab}$ in Ref.\ \cite{Ashtekar-Das}, we will work with $C_{abcd}$ as we described in Introduction.}.

A remarkable consequence that follows from this definition is that we obtain a conservation law associated with the asymptotic symmetry group at ${\cal I}$, which is found to be the $n$-dimensional AdS group $O(n-1,2)$.
The vectors $\xi^a$ that generate this asymptotic symmetry group are shown to be conformal Killing vectors at ${\cal I}$, and the conservation law at ${\cal I}$ is described as
\begin{equation}
\hat{D}^a (\hat{\cal E}_{ab} \xi^b) = - 8\pi G_{(n)} (n-3) \tau_{ab} \hat{n}^a \xi^b,
\label{conservation_Egrav}
\end{equation}
where $\hat{\cal E}_{ab}$ is the ``electric part'' of the Weyl tensor defined by
\begin{equation}
\hat{\cal E}_{a b} = \ell^2 K_{a c b d} \hat{n}^c \hat{n}^d
\label{eqn:EleWeylDef}
\end{equation}
and $\hat{n}_a$ is a normal vector to ${\cal I}$ defined as $\hat{n}_a \equiv \hat{\nabla}_a \Omega$ with its magnitude at ${\cal I}$ equal to the inverse of the curvature length $\ell$ as $\ell^{- 1} \equiv \sqrt{\hat{n}^a \hat{n}_a}$.
The induced metric $\hat{h}_{ab}$ on ${\cal I}$ is thus defined by $\hat{h}_{ab} \equiv \hat{g}_{ab} - \ell^2 \hat{n}_a \hat{n}_b$, and $\hat{D}_a$ denotes the covariant derivative associated with $\hat{h}_{ab}$.
Then, by integrating Eq.\ (\ref{conservation_Egrav}) over a portion $\Delta {\cal I}$ of ${\cal I}$, it is shown that there exist well-defined conserved quantities $Q_{\xi}[C]$ associated with the conformal Killing vectors $\xi^a$ at ${\cal I}$, which are defined on an $(n - 2)$-dimensional spacelike cross section $C$ of ${\cal I}$ as
\begin{equation}
Q_{\xi}[C] = - \frac{\ell}{8\pi G_{(n)}(n-3)} \int_C dx^{n-2} \sqrt{\hat{\sigma}} \hat{\cal E}_{ab} \xi^a \hat{N}^b,
\label{conservedquantitiesE}
\end{equation}
where $\hat{N}^b$ is the timelike unit normal to $C$, and $\hat{\sigma}$ is the determinant of the induced metric on $C$.
From the conservation law Eq.\ (\ref{conservation_Egrav}), these conserved quantities $Q_{\xi}[C]$ are found to satisfy the balance equation of the form
\begin{equation}
Q_{\xi}[C_2] - Q_{\xi}[C_1] = \ell \int_{\Delta {\cal I}}dx^{n-1}\sqrt{-\hat{h}}\tau_{ab}\xi^a \hat{n}^b,
\label{balanceeq}
\end{equation}
where $C_1$ and $C_2$ are the past and future boundaries of $\Delta {\cal I}$, respectively.
This indicates that decrease of a conserved quantity coincides with the corresponding flux of matter fields carried away across ${\cal I}$, and hence there is no unphysical increase or decrease of these conserved quantities.
(No gravitational radiation is allowed, and hence these conserved quantities are {\it absolutely conserved} when matter flux is absent \cite{Ashtekar-Magnon,Ashtekar-Das}.)

Here we note that the fall-off condition on the Weyl tensor can be weakened, as it is described in Ref.\ \cite{Ashtekar-Das}.
Since the conserved quantities $Q_{\xi}[C]$ are constructed only from the electric part of the Weyl tensor, the fall-off condition on other components (the magnetic part) is not necessary as long as the conserved quantities are concerned.
However, it is not certain whether a weakened fall-off condition makes any sense, and the definition in Einstein gravity formulated in Ref.\ \cite{Ashtekar-Das} is well-established.
Thus, in this paper, we will respect the definition in Ref.\ \cite{Ashtekar-Das}, and extend it to the case of higher curvature gravitational theories, where all components of the Weyl tensor are treated on an equal footing.

\subsection{Higher curvature gravitational theories}

In the case of Einstein gravity with a negative cosmological constant, the Einstein equation and the fall-off condition on the energy-momentum tensor $T_{ab}$ of matter fields determine the asymptotic form of the Ricci tensor, i.e., the Ricci tensor $R_{ab}$ associated with the physical metric $g_{a b}$ approaches its asymptotic value as
\begin{equation}
R_{ab} \rightarrow - \frac{n-1}{\ell^{2}} g_{ab},
\label{eqn:RicciAsym}
\end{equation} 
at the rate of $\Omega^{n-2}$.
Since the fall-off condition on the Weyl tensor also is imposed, the asymptotic behavior of all components of the curvatures tensors in Einstein gravity is specified by the conditions for an asymptotically AdS spacetime.
However, it is not always the case when higher curvature corrections are included, since the gravitational field equation is nonlinear in the curvature tensors and includes derivatives of the curvature tensors.
A spacetime may fail to approach the pure AdS spacetime at infinity in higher curvature gravitational theories, even if it satisfies all the conditions for an asymptotically AdS spacetime imposed in Einstein gravity.
Indeed, as we see from the results in Ref.\ \cite{EGBBH}, Einstein-Gauss-Bonnet gravity {\it with a negative cosmological 
constant} has a solution that asymptotically approaches the Minkowski or the de Sitter spacetime.
(See also the argument in Sec.\ \ref{subsec:Comparison} below.)

It is therefore obvious that one needs to supplement a condition that ensures the desired asymptotic form of the Ricci tensor as described by Eq.\ (\ref{eqn:RicciAsym}).
Not only the condition on the Weyl tensor, but also a condition on the Ricci tensor is necessary in higher curvature gravitational theories.
In addition, both the Ricci tensor and the Weyl tensor enter the gravitational field equation, in contrast to Einstein gravity.
It is then natural to consider all components of the curvature tensors on an equal footing, not considering separately the Weyl tensor and the Ricci tensor, as the electric part and the magnetic part of the Weyl tensor are treated on an equal footing in Einstein gravity.
Therefore, here we impose the condition that the Riemann tensor $R_{abcd}$ associated with the physical metric $g_{a b}$ should behave asymptotically as
\begin{equation}
R_{abcd} \rightarrow -\frac{2}{\ell^2}g_{c[a}g_{b]d},
\label{condi_Riemann_AdS}
\end{equation}
instead of imposing the condition that the Weyl tensor associated with $\hat{g}_{ab}$ should vanish at infinity.
Furthermore, since the Weyl tensor $C_{abcd}$ associated with $g_{ab}$ should fall off as $\Omega^{n-5}$ in Einstein gravity, we require also that the Riemann tensor $R_{abcd}$ should approach the form of Eq.\ (\ref{condi_Riemann_AdS}) at the rate of $\Omega^{n-5}$.

Thus, we now propose a definition of an asymptotically AdS spacetime compatible with higher curvature corrections, as follows.
\begin{definition}
An $n$-dimensional spacetime $(M, g_{ab})$ is said to be an asymptotically AdS spacetime if there exists a spacetime $(\hat{M},\hat{g}_{ab})$ equipped with the smooth metric $\hat{g}_{ab}$ and the boundary ${\cal I}$, which satisfies following conditions:
\begin{enumerate}
\item $\hat{M}\setminus{\cal I}\cong M$ and ${\cal I}\cong S^{n-2}\times {\bf R}$.
\item There exists a smooth scalar $\Omega$ on $\hat{M}$ such that
\begin{eqnarray}
\hat{g}_{ab}&=&\Omega^2 g_{ab}~~{\rm on}~~M,\\
\Omega&=&0~~{\rm on}~~{\cal I},\\
\hat{\nabla}_a\Omega&\ne&0~~{\rm on}~~{\cal I}.
\end{eqnarray}
\item The physical metric $g_{ab}$ satisfies the gravitational equation on $M$, and there exists a smooth tensor $\tau_{ab}$ such that the energy-momentum tensor $T_{ab}$ in a neighborhood of ${\cal I}$ is given by
\begin{equation}
T_{ab} = \Omega^{n-2} \tau_{ab}.
\label{Tab_falloff}
\end{equation}
\item There exist a smooth tensor $H_{abcd}$ and a real constant $\ell$ such that the Riemann tensor $R_{abcd}$ associated with the physical metric $g_{ab}$ in a neighborhood of ${\cal I}$ is given by
\begin{equation} 
R_{abcd} + \frac{2}{\ell^2}g_{c[a}g_{b]d} = \Omega^{n-5} H_{abcd}.
\label{Riemann_falloff}
\end{equation}
\end{enumerate}
\end{definition}

Thus, the first three conditions remain the same as in the case of Einstein gravity \cite{Ashtekar-Magnon,Ashtekar-Das}, while the condition on the Weyl tensor in the case of Einstein gravity is now replaced by the fourth condition on the Riemann tensor.
For a practical reason, however, it is convenient to decompose the Riemann tensor $R_{abcd}$ into the Ricci scalar $R$, the traceless part of the Ricci tensor $r_{ab}$ defined by
\begin{equation}
r_{ab} \equiv R_{ab} - \frac{1}{n} g_{ab} R,
\end{equation}
and the Weyl tensor $C_{abcd}$.
Then, the asymptotic condition on the Riemann tensor Eq.\ (\ref{Riemann_falloff}) is described as
\begin{eqnarray}
R &=& - \frac{n ( n - 1 )}{\ell^2} + \Omega^{n-1} N,
\label{R_falloff}\\
r_{ab} &=& \Omega^{n-3} L_{ab},
\label{rab_falloff}\\
C_{abcd} &=& \Omega^{n-5} K_{abcd},
\label{Weyl_falloff} 
\end{eqnarray}
where $N$, $L_{ab}$, and $K_{abcd}$ are smooth tensors.
In what follows, we will thus employ Eqs.\ (\ref{R_falloff})--(\ref{Weyl_falloff}) and denote the value of the Ricci scalar at ${\cal I}$ as
\begin{equation}
R_0 \equiv - \frac{n ( n - 1 )}{\ell^2}.
\label{eqn:RicciScalarInf}
\end{equation}
We note that the fall-off condition on the Weyl tensor remains the same as in Einstein gravity.
On the other hand, the Ricci tensor is required to approach its asymptotic value at the rate of $\Omega^{n-3}$.
We thus see that the asymptotic condition on the Ricci tensor required by the above definition is {\it weaker} than that required by the definition in Einstein gravity.

\subsection{Structure and fields at infinity}

From Eqs.\ (\ref{R_falloff})--(\ref{Weyl_falloff}), and the transformation law of the curvature tensors under conformal transformations, the asymptotic behavior of the Ricci scalar $\hat{R}$, the traceless part of the Ricci tensor $\hat{r}_{ab}$, and the Weyl tensors $\hat{C}_{abcd}$ associated with the unphysical metric $\hat{g}_{ab}$ is found as
\begin{eqnarray} 
\hat{R} &=& n (n-1) \Omega^{-2} \left[-\frac{1}{\ell^2} + \hat{n}^a \hat{n}_a \right]
- 2 \Omega^{-1} (n-1) \hat{\dalm} \Omega + \Omega^{n-3} N,
\label{R_falloff_2}\\
\hat{r}_{ab} &=& \Omega^{n-3} L_{ab}
-\Omega^{-1} (n-2) \left[\hat{\nabla}_a \hat{\nabla}_b \Omega - \frac{1}{n} \hat{g}_{ab} \hat{\dalm} \Omega \right],
\label{rab_falloff_2}\\
\hat{C}_{abcd} &=& \Omega^{n-3} K_{abcd},
\label{Weyl_falloff_2} 
\end{eqnarray}
respectively, where $\hat{\dalm}$ is the d'Alembertian associated with $\hat{g}_{ab}$ and $\hat{n}_a \equiv \hat{\nabla}_a \Omega$.
Since $\hat{g}_{ab}$ and hence $\hat{R}$ are smooth on ${\cal I}$, we see, by taking the limit to ${\cal I}$ of Eq.\ (\ref{R_falloff_2}), that $\hat{n}_a$ is spacelike at ${\cal I}$ as
\begin{equation}
\hat{n}^a \hat{n}_a = \frac{1}{\ell^2},
\label{n_R_K_relation}
\end{equation}
and hence that ${\cal I}$ is a timelike hypersurface.
Actually, $\hat{n}^a \hat{n}_a$ is positive in a neighborhood of ${\cal I}$, and thus the hypersurfaces $\Omega = {\it const.}$ are timelike not only right on ${\cal I}$, but also in a neighborhood of ${\cal I}$.
It is also shown, without using the field equation, that the asymptotic symmetry group at ${\cal I}$ is untouched by higher curvature corrections.
Indeed, by taking the limit to ${\cal I}$ of Eq.\ (\ref{rab_falloff_2}), we obtain
\begin{equation}
\hat{\nabla}_a\hat{\nabla}_b\Omega=\frac{1}{n}\hat{g}_{ab}\hat{\dalm}\Omega
\label{eqn:UnCondGauge}
\end{equation}
at ${\cal I}$.
As hown in Ref.\ \cite{Ashtekar-Das}, there exists conformal completion that satisfies
\begin{equation}
\hat{\nabla}_a\hat{\nabla}_b\Omega=0,
\label{Kab_atI}
\end{equation}
which implies that the extrinsic curvature of ${\cal I}$ vanishes.
Furthermore, there still remains conformal freedom that leaves Eq.\ (\ref{Kab_atI}) unaltered.
It then follows from this fact and Eq.\ (\ref{Weyl_falloff_2}), along with the `reflective boundary condition' in four dimensions, that ${\cal I}$ is conformally flat and hence that the asymptotic symmetry group is the $n$-dimensional AdS group $O(n-1,2)$.

To construct the conserved quantities associated with the generating vectors $\xi^a$ of the asymptotic symmetry group at ${\cal I}$, we employ the gravitational field equation, which we write as
\begin{equation}
E^{(g)}_{ab}=8\pi G_{(n)}T_{ab},
\label{G-eq}
\end{equation}
and define a tensor $P_{ab}$ by 
\begin{equation}
P_{ab}\equiv E^{(g)}_{ab}-\frac{1}{n-1}g_{ab}E^{(g)}_{cd}g^{cd}.
\label{P_ab_definition}
\end{equation}
From Eqs.\ (\ref{G-eq}) and (\ref{P_ab_definition}), we then obtain 
\begin{equation}
\nabla_{[e}P_{a]b} = 8\pi G_{(n)}\left[ \nabla_{[e}T_{a]b}-\frac{1}{n-1}\hat{g}_{b[a}\nabla_{e]}T_c{}^c\right],
\label{eqn:DerP}
\end{equation}
where $\nabla_a$ is the covariant derivative associated with the physical metric $g_{ab}$.
By using the fall-off condition on the energy-momentum tensor Eq.\ (\ref{Tab_falloff}) and the relation between the derivative operators $\hat{\nabla}_a$ and $\nabla_a$, we see that Eq.\ (\ref{eqn:DerP}) is written in a neighborhood of ${\cal I}$ as
\begin{equation}
\nabla_{[e}P_{a]b} = 8\pi G_{(n)} \Omega^{n-3} \biggl[ -\hat{n}^d\hat{g}_{b[e}\tau_{a]d} +(n-1)\hat{n}_{[e}\tau_{a]b} -\frac{n}{n-1}\hat{n}_{[e}\hat{g}_{a]b}\tau_c{}^c\biggr] + O(\Omega^{n-2}).
\label{conservationlaw_1}
\end{equation}
We multiply Eq.\ (\ref{conservationlaw_1}) by $\Omega^{-(n-3)} \hat{n}^e \hat{n}^b \hat{h}^a{}_c$, where $\hat{h}_{ab}$ is the induced metric on a hypersurface $\Omega = {\it const.}$, take the limit to ${\cal I}$, and then multiply it by $\xi^c$.
It yields
\begin{equation}
\Omega^{-(n-3)} (\nabla_{[e}P_{a]b}) \hat{n}^e \hat{n}^b \xi^a = \frac{n-2}{2\ell^2} 8 \pi G_{(n)} \tau_{ab} \hat{n}^b \xi^a.
\label{conservationlaw_2}
\end{equation}
If the left-hand side of Eq.\ (\ref{conservationlaw_2}) is rewritten into a form of divergence of a certain vector at ${\cal I}$, Eq.\ (\ref{conservationlaw_2}) gives a conservation law, like Eq.\ (\ref{conservation_Egrav}) in Einstein gravity, and thus we can define conserved quantities $Q_{\xi}[C]$ that satisfy the balance equation (\ref{balanceeq}).
We will see in the next section that this is indeed the case at least in two classes of higher curvature gravitational theories.

\section{Conserved quantities}
\label{sec:ConservedQuantities}

Based on the definition of an asymptotically AdS spacetime in the previous section, we now derive a conservation law at ${\cal I}$ and construct well-defined conserved quantities in higher curvature gravitational theories.
However, it is not straightforward to prove that the left-hand side of Eq.\ (\ref{conservationlaw_2}) can be generally rewritten into a form of divergence, since the tensor $P_{a b}$ depends strongly on the Lagrangian under consideration.
Therefore, we will focus in this section on two classes of higher curvature gravitational theories and see explicitly that Eq.\ (\ref{conservationlaw_2}) indeed gives a conservation law and hence conserved quantities.
We will also compare the conserved quantities constructed here with those derived elsewhere.

\subsection{$f(R)$-gravity}

We first consider the class of gravitational theories whose gravitational Lagrangian ${\cal L}(g^{ab}, R_{abcd})$ is given by a smooth function $f(R)$ of the Ricci scalar $R$ alone as
\begin{equation}
{\cal L}(g^{ab}, R_{abcd}) = f(R),
\label{fR_lagrangian}
\end{equation}
which we call $f(R)$-gravity in this paper.
The gravitational field equation in this class of gravitational theories is derived as
\begin{equation}
f'(R)R_{ab}-\frac{1}{2}g_{ab}f(R)+g_{ab}\dalm f'(R) -\nabla_a\nabla_b f'(R) = 8\pi G_{(n)}T_{ab},
\label{fR_GeqO}
\end{equation}
which contains fourth derivatives of the metric $g_{a b}$, when $f(R)$ is nonlinear in $R$.
Hence, $f(R)$-gravity possesses an extra degree of freedom due to the higher curvature corrections.

From Eq.\ (\ref{fR_GeqO}), we obtain $P_{ab}$ in $f(R)$-gravity as
\begin{equation}
P_{ab} = f'(R)r_{ab}-\frac{1}{n(n-1)}g_{ab}f'(R)R
+\frac{1}{2(n-1)}g_{ab}f(R)-\nabla_a\nabla_bf'(R).
\label{eqn:PDeffR}
\end{equation}
We then take the derivative of Eq.\ (\ref{eqn:PDeffR}), antisymmetrize the first two indices, and rewrite third derivatives of $R$ as
\begin{equation}
\nabla_{[e}\nabla_{a]}\nabla_b R =
\frac{1}{2} R_{eabd} g^{cd} \nabla_c R
= - \frac{1}{n(n-1)} R ( \nabla_{[e} R) g_{a]b} + O(\Omega^{2n-5}),
\label{del_boxabR}
\end{equation}
where we note that nonlinear terms including $r_{ab}$ or $C_{abcd}$ fall off faster than the first term in the right-hand side of Eq.\ (\ref{del_boxabR}).
In order to convert $\nabla_{[e}P_{a]b}$ into the desired form, we also apply the contracted Bianchi identity
\begin{equation}
\frac{2(n-3)}{n-2} \nabla_{[e} r_{a]b} 
+ \frac{n-3}{n(n-1)} ( \nabla_{[e} R) g_{a]b} 
= - g^{cd} \nabla_c C_{eabd},
\label{Bianchi_1}
\end{equation}
as well as the relation
\begin{equation}
g^{cd}\nabla_c C_{eabd}=\Omega^{n-3}\hat{\nabla}^c K_{eabc},
\label{DC_DK_relation}
\end{equation}
which we obtain by differentiating Eq.\ (\ref{Weyl_falloff_2}) and rewriting the derivative operator $\nabla_a$ in terms of $\hat{\nabla}_a$.
By imposing the asymptotic conditions Eqs.\ (\ref{R_falloff}) and (\ref{rab_falloff}), we eventually find that $\nabla_{[e}P_{a]b}$ is written as
\begin{equation}
\nabla_{[e}P_{a]b} = -\Omega^{n-3}\frac{n-2}{2(n-3)} f'(R_0) \hat{\nabla}^c K_{eabc} + O(\Omega^{2n-5}),
\label{DP_fR_2}
\end{equation}
where $R_0$ is defined by Eq.\ (\ref{eqn:RicciScalarInf}).
Notice that the second term in the right-hand side of Eq.\ (\ref{DP_fR_2}) falls off faster than the first term, and hence it does not make a contribution in Eq.\ (\ref{conservationlaw_2}).
From Eqs.\ (\ref{conservationlaw_2}) and (\ref{DP_fR_2}), we indeed obtain a conservation law at ${\cal I}$ as
\begin{equation}
f'(R_0) \hat{D}^c ( \hat{\cal E}_{c d} \xi^d )
= - 8\pi G_{(n)} (n-3) \tau_{ab} \hat{n}^a \xi^b,
\label{conservationfR}
\end{equation}
where $\hat{\cal E}_{a b}$ is defined by Eq.\ (\ref{eqn:EleWeylDef}).
Hence the conserved quantities $Q_{\xi}[C]$ are well-defined as
\begin{equation}
Q_{\xi}[C] = - \frac{f'(R_0) \, \ell}{8\pi G_{(n)}(n-3)} \int_C dx^{n-2} \sqrt{\hat{\sigma}} \hat{\cal E}_{ab} \xi^a \hat{N}^b,
\label{conserved_quantitiesfR}
\end{equation}
and satisfy the balance equation Eq.\ (\ref{balanceeq}) as in the case of Einstein gravity \cite{Ashtekar-Magnon,Ashtekar-Das}.
We note that the conserved quantities $Q_{\xi}[C]$ given by Eq.\ (\ref{conserved_quantitiesfR}) are expressed as an integral of the electric part of the Weyl tensor, while the proportion factor $f'(R_0)$ obviously depends on the Lagrangian.

\subsection{Quadratic Curvature Gravity}

Secondly, we consider quadratic curvature gravity, where the gravitational Lagrangian ${\cal L}(g^{ab}, R_{abcd})$ is given by
\begin{equation}
{\cal L}(g^{ab}, R_{abcd}) = - 2 \Lambda + R + \alpha {\cal L}_{\rm GB} + \beta R^2 + \gamma R_{ab} R^{ab}.
\label{GB_lagrangean}
\end{equation}
Here the constants $\alpha$, $\beta$, and $\gamma$ are coupling constants of higher curvature corrections, and ${\cal L}_{\rm GB}$ is the Gauss-Bonnet combination of the curvature tensors defined by
\begin{equation}
{\cal L}_{\rm GB} = R^2 - 4R_{ab} R^{ab} + R_{abcd} R^{abcd}.
\end{equation}
In this class of gravitational theories, $E^{(g)}_{ab}$ is derived as
\begin{eqnarray}
E^{(g)}_{ab} & = & g_{ab}\Lambda + R_{ab} - \frac{1}{2}g_{ab}R + 
2\alpha \biggl[R_{ab}R-2R_{ac}{R^c}_b 
- 2R_{acbd}R^{cd} + {R_a}^{cde}R_{bcde} 
- \frac{1}{4}g_{ab}{\cal L}_{\rm GB}\biggr] 
\nonumber\\ & & {} 
+ 2\beta \biggl[R_{ab}R - \frac{1}{4}g_{ab}R^2\biggr] 
+ 2\gamma \biggl[R_{acbd}R^{cd} 
- \frac{1}{4}g_{ab}R_{cd}R^{cd}\biggr] 
\nonumber \\ & & {} 
+ (2\beta+\gamma)[g_{ab}\dalm-\nabla_a\nabla_b] R 
+ \gamma\dalm \biggl[ R_{ab} - \frac{1}{2}g_{ab}R \biggr] ,  
\label{G-eq_quadraticgrav}
\end{eqnarray}
and hence the field equation contains fourth derivatives of the metric in general, while the theory reduces to Einstein-Gauss-Bonnet gravity when $\beta=\gamma=0$, where the gravitational equation includes the second derivatives at most \cite{EGB}.

From Eq.\ (\ref{G-eq_quadraticgrav}), $P_{ab}$ in quadratic curvature gravity is calculated as
\begin{equation}
P_{ab} = - \frac{1}{n-1}g_{ab}\Lambda + r_{ab} + 
\frac{n-2}{2n(n-1)}g_{ab}R 
+\alpha P^{(\alpha)}_{ab} + \beta P^{(\beta)}_{ab} + \gamma P^{(\gamma)}_{ab},
\label{fR_QCG}
\end{equation}
where
\begin{eqnarray}
P^{(\alpha)}_{ab} &=& \frac{2(n-3)(n-4)}{n(n-1)}Rr_{ab} 
+ \frac{(n-2)(n-3)(n-4)}{2n^2(n-1)^2}g_{ab}R^2 
- \frac{4(n-3)(n-4)}{(n-2)^2}r_{ac}r^c{}_b 
\nonumber\\
& &{} 
+ \frac{2(n-3)(n-4)}{(n-1)(n-2)^2}g_{ab}r_{cd}r^{cd} 
- \frac{4(n-4)}{n-2}C_{acbd}r^{cd}+2C_{acdf}C_b{}^{cdf} 
-\frac{3}{2(n-1)}g_{ab}C_{cdfg}C^{cdfg} , 
\label{Pab_alpha}\\
P^{(\beta)}_{ab} &=& 2Rr_{ab} + \frac{n-4}{2n(n-1)}g_{ab}R^2 - 2\nabla_a\nabla_b R , 
\label{Pab_beta}\\
P^{(\gamma)}_{ab} &=& \frac{3n-4}{n(n-1)}Rr_{ab} 
+ \frac{n-4}{2n^2(n-1)}g_{ab}R^2 
+\frac{n-4}{n-2}r_{ac}r^c{}_b - \frac{n-4}{2(n-1)(n-2)}g_{ab}r_{cd}r^{cd} 
\nonumber\\
& &{} 
+C_{acbd}r^{cd} - \frac{n}{2(n-1)}\nabla_a\nabla_bR 
+\frac{n-2}{n-3}\nabla^c\nabla^d C_{acbd} . 
\label{Pab_gamma}
\end{eqnarray}
Note that in Eqs.\ (\ref{GB_lagrangean})--(\ref{Pab_gamma}), 
the tensor indices are raised by the physical metric $g^{ab}$.

As in the previous subsection, we compute $\nabla_{[e}P_{a]b}$, by imposing the asymptotic condition Eqs.\ (\ref{R_falloff})--(\ref{Weyl_falloff}), and using Eqs.\ (\ref{del_boxabR})--(\ref{DC_DK_relation}).
Nonlinear terms in $r_{ab}$ or $C_{abcd}$ are then found to fall off fast enough, and we have
\begin{eqnarray}
\nabla_{[e}P_{a]b} & = & \Omega^{n-3}\biggl\{\biggl[-\frac{n-2}{2(n-3)} 
+\frac{1}{\ell^2}\biggl(\alpha(n-2)(n-4) 
+\beta\frac{n(n-1)(n-2)}{(n-3)} 
+\gamma\frac{(n-2)(3n-4)}{2(n-3)}\biggr)\biggr] 
\hat{\nabla}^c K_{eabc} 
\nonumber\\
& &{} 
+\gamma\frac{(n-2)(n-4)}{n-3}\Bigl[(n-3)\hat{n}^c\hat{n}^d\hat{\nabla}_{[e}K_{a]cbd} 
-2\hat{n}^c\hat{n}_{[e}\hat{\nabla}^d K_{a]cbd}-2\hat{n}^c\hat{n}_{[e}\hat{\nabla}^d K_{a]dbc} 
-\hat{n}^c\hat{n}^d\hat{g}_{b[e}\hat{\nabla}^f K_{a]cfd}\Bigr]\biggr\} 
\nonumber\\
& & {} + O(\Omega^{2n-5}).
\label{DP_QCG_2}
\end{eqnarray}
{}From Eqs.\ (\ref{conservationlaw_2}) and (\ref{DP_QCG_2}), we thus obtain a conservation law at ${\cal I}$ as
\begin{equation}
\Xi_q \hat{D}^c ( \hat{\cal E}_{cd} \xi^d ) 
= - 8\pi G_{(n)} (n-3) \tau_{ab} \hat{n}^a \xi^b , 
\label{conservationq}
\end{equation} 
where $\hat{\cal E}_{a b}$ is defined by Eq.\ (\ref{eqn:EleWeylDef}), $\Xi_q$ is given as
\begin{equation}
\Xi_q = 1 + R_0 \left[ 2 \alpha \frac{(n-3)(n-4)}{n ( n - 1 )} 
+ 2 \beta +4 \gamma \frac{(n-2)}{n ( n - 1)} \right] , 
\label{eqn:PropFactQdr} 
\end{equation}
and $R_0$ is defined by Eq.\ (\ref{eqn:RicciScalarInf}).
Accordingly, we can construct the well-defined conserved quantities $Q_{\xi}[C]$ as
\begin{equation}
Q_{\xi}[C] = - \frac{\Xi_q \ell}{8\pi G_{(n)}(n-3)} \int_C dx^{n-2} \sqrt{\hat{\sigma}} \hat{\cal E}_{ab} \xi^a \hat{N}^b,
\label{conserved_quantitiesq}
\end{equation}
which satisfy the balance equation Eq.\ (\ref{balanceeq}).
Again, we see that the conserved quantities $Q_{\xi}[C]$ are expressed as an integral of the electric part of the Weyl tensor, while it is multiplied by the proportion factor $\Xi_q$ that depends on the coupling constants $\alpha$, $\beta$, and $\gamma$, as well as the cosmological constant $\Lambda$.

\subsection{Comparison}
\label{subsec:Comparison}

In order to confirm that the results in this paper are reasonable, it is crucial to see that the expressions of the conserved quantities we derived above are consistent with those derived in other formalisms.
In particular, black hole solutions in Einstein-Gauss-Bonnet gravity (quadratic curvature gravity with $\beta=\gamma=0$) have been investigated so far.
Thus, here we compare Eq.\ (\ref{conserved_quantitiesq}) with the results in Einstein-Gauss-Bonnet gravity established elsewhere.

A higher-dimensional ($n \geq 5$) spherically symmetric black hole solution in Einstein-Gauss-Bonnet gravity has been derived \cite{EGBBH}, which describes an asymptotically AdS black hole spacetime for suitable combinations of $\Lambda$ and $\alpha$.
The metric of this solution is given by
\begin{equation}
ds^2 = - f(r) dt^2 + f^{-1}(r) dr^2 + r^2 d \sigma_{n-2}^2,
\label{EGBSphSol} 
\end{equation}
where $f(r)$ is written as
\begin{equation}
f(r) = 1 + \frac{r^2}{2 (n-3) (n-4) \alpha} \left[ 1 \pm 
\sqrt{1 + \frac{64\pi G_{(n)} (n-3) (n-4) \alpha {\cal M}}{(n-2) V_{n-2} r^{n-1}} 
+ \frac{8 (n-3) (n-4) \alpha \Lambda}{(n-1) (n-2)} }\right],
\label{eqn:MetricFuncCai}
\end{equation}
$d\sigma_{n-2}^2$ is the metric of an $(n-2)$-dimensional unit sphere, $V_{n-2}$ is its volume, and it has been shown that the mass of this black hole is given by the constant ${\cal M}$ \cite{EGBBH,DeruelleKatzOgushi,energyEGBBH}.

As we see from Eq.\ (\ref{eqn:MetricFuncCai}), there exist two branches of black hole solutions corresponding to the signs in front of the square root.
The solutions with the lower sign reduce to solutions in Einstein gravity in the limit of $\alpha \rightarrow 0$, while the solutions with the upper sign do not.
The branch of the upper sign arises essentially because the field equation is nonlinear in the curvature tensors.
To see this in the context of this paper, we consider the trace of the field equation $E^{(g)}_{ab} = 0$, where $E^{(g)}_{ab}$ is given by Eq.\ (\ref{G-eq_quadraticgrav}) with $\beta = \gamma = 0$, and take the limit to ${\cal I}$.
It then yields
\begin{equation}
- \frac{n - 2}{2} R_0 + n \Lambda 
- \alpha \frac{( n - 2 ) ( n - 3 ) ( n - 4 )}{2 n ( n - 1)} R_0^2 
= 0
\label{R_0_condi_QCG}
\end{equation}
at ${\cal I}$, where $R_0$ is the value of the Ricci scalar at ${\cal I}$.
By solving Eq.\ (\ref{R_0_condi_QCG}) for $R_0$, we obtain
\begin{equation}
1 + R_0 \frac{2 ( n - 3 ) ( n - 4 ) \alpha}{n ( n - 1 )} = 
\mp \sqrt{1 + \frac{8 (n-3) (n-4) \alpha \Lambda}{(n-1) (n-2)}} , 
\label{ellvalue_QCG}
\end{equation}
which gives, by using Eq.\ (\ref{eqn:RicciScalarInf}),
\begin{equation} 
\frac{1}{\ell^2} = \frac{1}{2 (n-3) (n-4) \alpha} \left[ 1 \pm 
\sqrt{1 + \frac{8 (n-3) (n-4) \alpha \Lambda}{(n-1) (n-2)}} \right].
\label{eqn:EllEGB}
\end{equation}
Therefore, two branches of real solutions are possible in Einstein-Gauss-Bonnet gravity, as long as the Ricci scalar converges to a constant.

It is straightforward to see that the asymptotic condition Eq.\ (\ref{Riemann_falloff}) on the Riemann tensor is indeed satisfied by both of the two branches of the solution Eq.\ (\ref{EGBSphSol}).
Hence we can apply Eq.\ (\ref{conserved_quantitiesq}) to compute the mass of this black hole solution by setting $\xi^a$ equal to the timelike Killing vector $( \partial / \partial t )^a$, as long as there exists a real solution of the curvature length $\ell$.
The timelike unit normal $\hat{N}^a$ (with respect to the unphysical metric $\hat{g}_{a b}$) to a cross section ${\cal C}$ of ${\cal I}$ is found to be given by $\hat{N}^a = \ell ( \partial / \partial t )^a$, where the curvature length $\ell$ is given as Eq.\ (\ref{eqn:EllEGB}) and it is understood that the conformal factor $\Omega$ is chosen here as $\Omega = 1 / r$.
We then see that only the $t$-$t$ component of $\hat{\cal E}_{ab}$ is involved in Eq.\ (\ref{conserved_quantitiesq}), which is calculated as
\begin{equation}
\hat{\cal E}_{tt} = -\frac{8\pi G_{(n)}(n-3)}{\Xi_q \ell^2 V_{n-2}}{\cal M}
\end{equation}
at ${\cal I}$ for both of the signs.
Therefore, we find that Eq.\ (\ref{conserved_quantitiesq}) yields $Q_{\xi}[C]={\cal M}$ for any cross sections $C$, and hence that Eq.\ (\ref{conserved_quantitiesq}) actually reproduces the result established in Ref.\ \cite{EGBBH,DeruelleKatzOgushi,energyEGBBH}.

Although very little has been known about rotating black hole solutions in Einstein-Gauss-Bonnet gravity, Deruelle and Morisawa \cite{DeruelleMorisawa} noticed that the $n$-dimensional Kerr-AdS spacetime \cite{GLPP} with its cosmological constant rescaled is an asymptotic solution at infinity even in Einstein-Gauss-Bonnet gravity.
Among the two branches of solutions in Einstein-Gauss-Bonnet gravity that we mentioned above, they thus focused on the branch corresponding to the lower sign in Eq.\ (\ref{ellvalue_QCG}).
Then they derived the expressions of the mass and the angular momenta of the Kerr-AdS spacetime in Einstein-Gauss-Bonnet gravity by using the KBL formalism \cite{DeruelleKatz,DeruelleKatzOgushi} and setting $\xi^a$ equal to the nonrotating timelike Killing vector at ${\cal I}$ \cite{GibbonsEtAl}.

When we compare the expressions derived in Ref.\ \cite{DeruelleMorisawa} with those in Einstein gravity \cite{GibbonsEtAl}, we find that the expressions in Einstein-Gauss-Bonnet gravity are given by those in Einstein gravity multiplied by the proportion factor given as
\footnote{Note, however, that it does not mean that the {\it values} of the conserved quantities in Einstein-Gauss-Bonnet gravity is equal to those in Einstein gravity multiplied by the factor $\Xi'$, since one needs to rescale the cosmological constant when one passes from Einstein gravity to Einstein-Gauss-Bonnet gravity.},
\begin{equation}
\Xi' = \sqrt{1 + \frac{8 (n-3) (n-4) \alpha \Lambda}{(n-1) (n-2)}}.
\label{eqn:PropFacDM}
\end{equation}
On the other hand, as we see from Eqs.\ (\ref{conservedquantitiesE}) and (\ref{conserved_quantitiesq}), the result in this paper also implies that the expression of conserved quantities in Einstein-Gauss-Bonnet gravity (in more general classes of gravitational theories, actually) is given by that in Einstein gravity multiplied by the proportion factor $\Xi_q$.
Furthermore, we see from Eqs.\ (\ref{eqn:PropFactQdr}), (\ref{ellvalue_QCG}), and (\ref{eqn:PropFacDM}) that $\Xi_q$ (with $\beta = \gamma = 0$) coincides with $\Xi'$ for the lower sign.
Thus, we immediately see that Eq.\ (\ref{conserved_quantitiesq}) reproduces the result in Ref. \cite{DeruelleMorisawa}, as well.

\section{Summary and discussion}
\label{sec:SummaryandDiscussion}

In this paper, we studied asymptotically AdS spacetimes in higher curvature gravitational theories by employing the conformal completion technique \cite{Ashtekar-Magnon,Ashtekar-Das}.
We first discussed that a condition on the Ricci tensor, which is automatically satisfied in Einstein gravity with a negative cosmological constant, should be supplemented in higher curvature gravitational theories, since otherwise a spacetime may fail to approach the pure AdS spacetime.
Based on the simple idea that all components of the Riemann tensor should be treated on an equal footing in higher curvature gravitational theories, we then proposed an alternative definition of an asymptotically AdS spacetime, where we imposed the condition on the Riemann tensor, instead of the Weyl tensor, as described by Eq. (\ref{Riemann_falloff}).

We then pointed out that the asymptotic symmetry group is unaffected by higher curvature corrections, and hence it is described by the $n$-dimensional AdS group $O(n-1, 2)$.
We explicitly showed for $f(R)$-gravity and quadratic curvature gravity that the definition proposed in this paper ensures that the conserved quantities associated with the asymptotic symmetry group are well-defined and satisfy the balance equation as in Einstein gravity.
These conserved quantities $Q_{\xi}[C]$ were shown to take the form
\begin{equation}
Q_{\xi}[C] = - \frac{\Xi \, \ell}{8\pi G_{(n)}(n-3)} \int_C dx^{n-2} \sqrt{\hat{\sigma}} 
\hat{\cal E}_{ab} \xi^a \hat{N}^b
\label{conserved_quantitiesUniv}
\end{equation}
in both $f(R)$-gravity and quadratic curvature gravity, where $\Xi$ is given by $f'(R_0)$ in $f(R)$-gravity and by $\Xi_q$ in quadratic curvature gravity.
Thus, $Q_{\xi}[C]$ are expressed as an integral of the electric part of the Weyl tensor alone, while it is multiplied by the proportion factor $\Xi$, which depends on the Lagrangian of a gravitational theory under consideration.
Therefore, in particular, all of the conserved quantities identically vanish in the pure AdS spacetime also in these higher curvature gravitational theories, as it is emphasized in the case of Einstein gravity in Ref.\ \cite{Ashtekar-Das}.

The Lagrangian of $f(R)$-gravity is an arbitrary function of 
the Ricci scalar $R$ alone, while the Lagrangian of quadratic curvature gravity includes all components of the curvature tensors but should be quadratic at most.
The results in these two classes of gravitational theories thus complement each other, based on which we can speculate on the behavior of conserved quantities in more general classes of gravitational theories.
By extrapolating the results in this paper, it is not difficult to expect that conserved quantities in an asymptotically AdS spacetime in an arbitrary gravitational theory are expressed as an integral of the electric part of the Weyl tensor alone.
Actually, nonlinear terms in $r_{a b}$ or $C_{a b c d}$ will not contribute in the left-hand side of Eq.\ (\ref{conservationlaw_2}), in general, because of the asymptotic condition on the curvature tensors.
Then, it is expected that remaining terms will be written in terms of the Weyl tensor alone, when we take the limit to ${\cal I}$ where the Ricci scalar converges to a constant, as in the cases we considered in this paper.
We have not proven in this paper whether it is indeed the case in general, but it will be interesting to investigate this issue further, since higher curvature corrections arise naturally in quantum gravity and conserved quantities in an asymptotically AdS spacetime play an important role in AdS / CFT correspondence.

We also showed that the expression of the conserved quantities constructed in this paper correctly reproduces the results derived elsewhere for particular black hole solutions in Einstein-Gauss-Bonnet gravity, when we set $\xi^a$ in Eq.\ (\ref{conserved_quantitiesq}) equal to the appropriate timelike Killing vector.
On the other hand, we have not explored the general connections between the conserved quantities in the conformal completion technique and those in other formalisms proposed so far \cite{Wald,Deser-Tekin}.
Although the conformal completion technique possesses the advantage that conserved quantities are defined intrinsically at infinity without referring to any background spacetimes, other formalisms also have their own virtues.
In particular, conserved quantities in the symplectic formalism \cite{Wald} are defined as the values of the canonical generators of an associated symmetry group.
Moreover, since the symplectic formalism provides the relation between conserved quantities defined at two distinct boundaries when there exist Killing vectors, they automatically satisfy the first law of black hole thermodynamics.
It is therefore important to clarify the general connections between conserved quantities in different formalisms, with which we will obtain deeper insights into physical aspects of conserved quantities in gravitational theories.

Finally we discuss an asymptotic condition on the curvature tensors.
Although we have shown that the asymptotic condition imposed in this paper is sufficient for the conserved quantities to be well-defined, we have not shown that it is necessary.
It looks reasonable to impose the condition on the electric part of the Weyl tensor, since otherwise the conserved quantities may diverge.
However, it is not straightforward to extrapolate it to the magnetic part of the Weyl tensor and even to the Ricci tensor.
Actually, even when the magnetic part of the Weyl tensor and the Ricci tensor approach their asymptotic values at ${\cal I}$ at the rates of $\Omega^{[ n / 2 ] - 3}$ and $\Omega^{[ n / 2 ] - 1}$ respectively, one can show that Eqs.\ (\ref{conservationfR}) and (\ref{conservationq}) remain valid, and hence the conserved quantities are well-defined in both $f(R)$-gravity and quadratic curvature gravity.
Furthermore, in four-dimensional $f(R)$-gravity, one can show that there exists a perturbative solution on the pure AdS background spacetime, where the asymptotic condition imposed in this paper is {\it not} satisfied but conserved quantities are actually well-defined.
From this point of view, the conditions imposed in this paper might seem too stringent.
However, recall here the fact that $f(R)$-gravity can be transformed into the Einstein frame, where the higher curvature corrections in $f(R)$-gravity are described by an extra scalar field.
Then, the energy-momentum tensor of the extra scalar field in the Einstein frame is found to fall off as $\Omega^{n - 3}$, 
which does not obey the condition for an asymptotically AdS spacetime in Einstein gravity.
In spite of this fact, one can show that the conserved quantities in this perturbative solution are well-defined also in the Einstein frame.
(Actually, one can show that the conserved quantities in the Einstein frame coincide with those in the original frame, as we will report elsewhere.)
Therefore, this perturbative solution, which might be considered as asymptotically AdS but fails to satisfy the asymptotic condition imposed in this paper, does not satisfy the conditions for an asymptotically AdS spacetime when viewed in the Einstein frame, either.
Hence, when we respect the definition in Ref.\ \cite{Ashtekar-Magnon,Ashtekar-Das}, the asymptotic condition imposed in this paper looks reasonable as long as this perturbative solution is concerned.
In addition, it seems sensible to impose an asymptotic condition on the Ricci tensor to control the asymptotic behavior of extra degrees of freedom that arise from higher curvature corrections.
However, in order to understand what is meant by an asymptotically AdS spacetime in higher curvature gravitational theories, refined analyses with regard to asymptotic conditions on the curvature tensors will be required.
Alternatively, we may replace the asymptotic conditions on the curvature tensors by other conditions.
For example, it will be possible to impose the condition that the conserved quantities associated with the asymptotic symmetry group should be well-defined.

\acknowledgments
We are grateful to N.\ Deruelle and Y.\ Morisawa for useful discussion and comments.
We are also grateful to K.\ Maeda for his continuous encouragement.
This work was partially supported by a Grant for The 21st Century COE Program (Holistic Research and Education Center for Physics Self-Organization Systems) at Waseda University.


\end{document}